\documentclass[%
 reprint,
superscriptaddress, prx, 
 amsmath,amssymb,
 aps,
]{revtex4-2}

\usepackage{graphicx}
\usepackage{dcolumn}
\usepackage{bm}
\usepackage{color}

\newcommand{\lac}{Universit\'e Paris-Saclay, CNRS, Laboratoire Aim\'e Cotton, 91405, Orsay, France.}
\newcommand{\uva}{Department of Physics, University of Virginia, Charlottesville, Virginia 22904, USA.}

\begin{document}


\title{A Coherent Light Shift on Alkaline-Earth Rydberg Atoms from Isolated Core Excitation without Auto-Ionization}

\author{Ky-Luc Pham}
\affiliation{\lac}
\author{Thomas Gallagher}
\email[Contact: ]{tfg@virginia.edu}
\affiliation{\uva}
\author{Pierre Pillet}
\affiliation{\lac}
\author{Steven Lepoutre}
\affiliation{\lac}
\author{Patrick Cheinet}
\email[Contact: ]{patrick.cheinet@u-psud.fr}
\affiliation{\lac}

\date{\today}

\begin{abstract}
New experimental quantum simulation platforms have recently been implemented with divalent atoms trapped in optical tweezer arrays with promising performance. The second valence electron also brings new propects through the so-called Isolated Core Excitation (ICE), however auto-ionization presents a strong limitation to this use. In this study, we propose and demonstrate a new approach to applying a sizable light shift to a Rydberg state with close-to-resonant ICE while avoiding auto-ionization. In particular, we have investigated ICE of ytterbium atoms in $^1S_0$ Rydberg states. Spectroscopic studies of the induced auto-ionization and the light shift imparted to the Rydberg states are perfectly accounted for with Multi-channel Quantum Defect Theory. Such a control over the inner electron without disturbing the Rydberg electron brings a new tool for the targeted, coherent manipulation of Rydberg states in quantum simulation experiments performed with alkaline-earth atoms.

\end{abstract}

\maketitle


\section{\label{sec:Introduction}Introduction}

Quantum simulation is an appealing prospect offered by experiments based on cold neutral atoms, and Rydberg excitations \cite{Gallagher1994} constitute an efficient tool to perform quantum operations \cite{Jaksch2000,Lukin2001,Weimer2010,Saffman2010}, providing strong interactions between electric dipoles \cite{Gallagher2008,Browaeys2016}. Experimental confirmation of these assets was first evidenced in quantum gas microscopes \cite{Schauss2012,Schauss2015,Zeiher2016,Gross2017}. Building on this progress, new platforms based on Rydberg excitations of alkali atoms trapped in arrays of optical tweezers were designed to engineer strongly correlated many-body ensembles \cite{Saffman2016,Browaeys2020,Henriet2020}. Groundbreaking experiments were reported with these platforms in the domains of quantum simulation \cite{Bernien2017,deLeseleuc2018a,Keesling2019,deLeseleuc2019,Schlosser2020,Ebadi2021,Scholl2021} and its digital counterpart, quantum computing \cite{Levine2018,Levine2019,Graham2019}, demonstrating the relevance of this approach for quantum information processing.

In the meantime, attention has been drawn to the opportunities offered by alkaline-earth species \cite{Mukherjee2011,Dunning2016}. The presence of a second valence electron provides a rich internal structure with attractive properties for quantum simulation \cite{Gorshkov2010,Gerbier2010} and quantum computing \cite{Daley2008,Gorshkov2009}, and the same goes for their Rydberg states \cite{Mukherjee2011,Dunning2016}. This structure also offers efficient laser cooling \cite{Stellmer2013}, leading to very effective simultaneous cooling and imaging in quantum gas microscopes \cite{Yamamoto2016} and optical tweezers \cite{Norcia2018,Cooper2018,Saskin2019,Covey2019}. It has led to a proof of principle demonstration of quantum operations with alkali-earth atoms with outstanding performance \cite{Madjarov2020}.

With alkali atoms, although some work-arounds have been found for undirect imaging \cite{Schauss2012,Gunter2013} and trapping in the expulsive ponderomotive potential on the Rydberg electron \cite{Dutta2000,Younge2010,Zhang2011,Anderson2011,Barredo2020,Cortinas2020}, the techniques to manipulate -- cool, image or light-shift -- ground state atoms become inefficient when applied to Rydberg atoms. A promising resource of alkaline-earth species resides in the Isolated Core Excitation (ICE) of a Rydberg state, which consists of addressing the optical transitions of the remaining valence core electron. Proof of principle experiments have demonstrated the interest of addressing specific sites with focused Rydberg excitation beams \cite{Graham2019} or with a coherent shift of the atom ground state \cite{Labuhn2014,Xia2015,Levine2019}, the latter with potentially detrimental effects on the gate fidelity due to alteration of the tweezer potential. Lack of control over the alkali Rydberg states expulsed by the tweezer light is also compelling \cite{Saffman2016,Levine2018}. The use of ICE instead should alleviate these issues.

However, a strong limitation in the application of ICE is the Auto-Ionization (AI) of a Rydberg state with an excited core, evidenced in historical pioneering experiments \cite{Cooke1978}. In this process, the electrostatic interaction between Rydberg and core electrons leads to the simultaneous ejection of one electron and relaxation of the other to a lower energy state. While AI can be turned into an asset as a means of diagnosis, as demonstrated on strontium \cite{Millen2011,McQuillen2013,Lochead2013}, it represents a major obstacle for the coherent, or at least lossless, manipulation of Rydberg states. Considering ICE far from resonance may allow its use as a trap \cite{Mukherjee2011,Wilson2019}, but residual losses are still constrained by AI rates which are usually much larger than radiative decay rates. Moreover, far off-resonant traps demand high laser power and do not address selectively Rydberg states, contrary to close-to-resonance ICE. Another way to mitigate AI rates is to use high orbital angular momentum ($\ell$ quantum number) Rydberg states \cite{Jones1988,Pruvost1991,Lehec2021}, called high-$\ell$ Rydberg states, which can have rates below the radiative decay rates for circular states \cite{Teixeira2020}. Consequently, ICE of circular states allows familiar approaches to cooling, imaging and coherent energy shift for dipole trapping, but at the price of increased experimental complexity. Here we introduce an ICE approach which allows large near resonant frequency shifts of low angular momentum Rydberg states while suppressing AI. It is based on the existence of minima of the AI rate occuring in the ICE spectra, previously observed with atomic beams \cite{Cooke1978,Safinya1979,Tran1982}, called auto-ionization zeros (AIz). They arise from the destructive interferences between multiple pathways leading to an auto-ionized state, through the Rydberg series with excited core.

Our team reported the first frozen Rydberg gas of ytterbium \cite{Lehec2018} excited from atoms in a Magneto-Optical Trap (MOT), and its quasi-resonant ICE near the first optical transition (wavelength 369.5 nm) of high-$\ell$ states \cite{Lehec2021}. Ytterbium is now used on another platform \cite{Saskin2019,Wilson2019} aiming at quantum computation based on trapped atoms. In this work, we report high resolution spectra for the AI of $6sns\,^1S_0$ states of ytterbium, and demonstrate the reduction of AI rates below our detection limit at the AIz. Eventually, we use Rydberg excitation spectroscopy to measure the induced ICE light shift as a function of detuning, and verify the existence of a sizeable shift at the AIz. All results are in good agreement with a Multichannel Quantum Defect Theory (MQDT) \cite{Seaton1983,Cooke1985} treatment of the effects of ICE light. While preparing this manuscript, we became aware of a similar study with $6sns\,^3S_1$ states of ytterbium which also uncovered useful AIz \cite{Burgers2021}.

\section{\label{sec:Experiment}Experimental protocol}

\begin{figure}[h]
	\includegraphics[width=0.9\columnwidth]{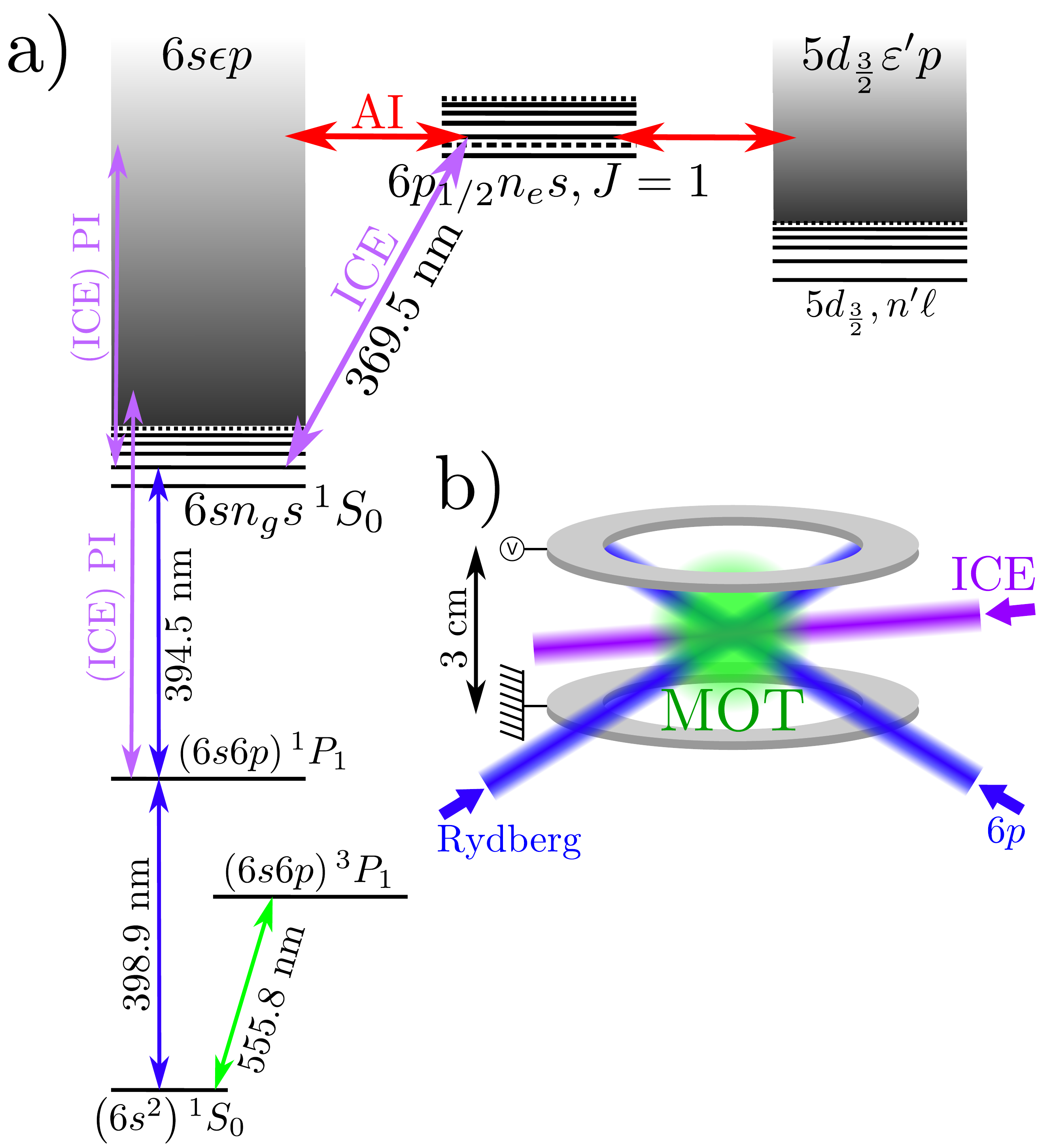}
	\caption{\label{fig:YbExperiment} a) Energy-level diagram of $^{174}$Yb related to the experiment and couplings between them. Purple arrows show couplings induced by the ICE light. It connects atoms in $6sn_gs\,^1S_0$ states to the $6p_{1/2}n_es, J=1$ Rydberg series. Red horizontal arrows show the coupling of this doubly excited series to the two continua $6s\epsilon p$ and $5d_{3/2}\epsilon^\prime p$, responsible for AI. Possible direct photo-ionization (PI) processes induced by ICE light are also shown (purple vertical arrows). b) Laser and electrode configuration around the MOT.}
\end{figure}

Fig. \ref{fig:YbExperiment} shows the experimental set-up designed to study the effects of ICE light on $^{174}$Yb Rydberg atoms. The atomic source is based on a MOT operated on the intercombination transition $\left(6s^2\right){^1S_0} \leftrightarrow \left(6s6p\right){^3P_1}$ at 555.8 nm, described in greater detail in \cite{Lehec2018}. It collects about $10^6$ atoms in a sphere of around 1 mm diameter. Three beams of approximatly horizontal propagation are shined on the ytterbium cloud after being passed through accousto-optic modulators to control their power. The measurement protocol consists of applying precisely timed sequences of pulses with durations in the $\mu$s range, repeated at a 10 Hz rate, with the MOT continuously loaded. The beams denoted $6p$ (wavelength 398.9 nm) and Rydberg (394.5 nm) provide two-photon transitions to reach $6sn_gs\,^1S_0$ Rydberg states (thereafter denoted $6sn_gs$), with $n_g=$ 50, 60 or 72. The ICE beam at wavelength 369.5 nm couples these Rydberg states to the doubly excited Rydberg series $6p_{1/2}n_es,J=1$ (denoted $6pn_es$), which can auto-ionize to the two continuum channels $6s\epsilon p$ and  $5d_{3/2}\epsilon^\prime p$ (red, horizontal arrows). As shown by the figure (purple, vertical arrows), direct Photo-Ionization (PI) by ICE light is also possible for the $6sn_gs$ and $\left(6s6p\right){^1P_1}$ states, with a much higher expected cross-section for the latter because the process ends up closer from the first ionization limit.

After each pulse sequence, an electric field ramp ionizes Rydberg states and sends all the ions towards a Multi-Channel Plate (MCP) for counting. Ions already present at the end of the pulse sequence are discriminated from ionized Rydberg state atoms by their time of flight. Gated integration of the MCP signals thus yields the number of ions $N_\text{Ion}$ and the number of Rydberg atoms $N_R$ at the end of the pulse sequence.

The 398.9 nm laser source frequency is locked by saturated spectroscopy on the $\left(6s^2\right){^1S_0} \leftrightarrow \left(6s6p\right){^1P_1}$ first optical transition of ytterbium, and the $6p$ beam frequency can be detuned from resonance by up to 1 GHz. The Rydberg and ICE laser light are produced by frequency-doubling two dedicated Ti:sapphire sources; the source frequencies are controlled by piezoelectric actuators, and measured by a wavemeter with a $\sim 1$ MHz short term precision and a $\sim 100$ MHz long term accuracy. The Rydberg frequency, denoted $f_{Ryd}$, is controlled by software means, while the ICE frequency $f_{ICE}$ is tuned manually in the vicinity of the $6s_{1/2} \leftrightarrow 6p_{1/2}$ ion transition frequency, $f_{ion} = \omega_{ion}/\left(2 \pi \right) = 811.291500(40)$ THz \cite{Zalivako2019}, and left free-running between manual corrections of its drift. The transverse sizes of the three beams at the atom cloud location, described by the waist radii $w_{6p}$, $w_{Ryd}$ and $w_{ICE}$, are adjusted using telescopes and/or focusing lenses prior to injection into the vacuum chamber. Time delays for their pulse application were precisely calibrated with a fast photodiode. Attention was also paid to the accurate definition and knowledge of the ICE beam properties. For mode-cleaning, it is first passed through a single mode fiber. $w_{ICE}$ was calibrated using a CCD camera after reproducing its optical path (including the chamber windows) on a table. The ICE light polarization is linear in the horizontal direction, and its total power $P_{ICE} \sim$ 20 - 50 mW was measured using a photodiode-based powermeter prior to each measurement.

To record AI spectra, we measure $N_\text{Ion}$ and $N_R$ as a function of $f_{ICE}$, typically scanned over several 100 GHz. We use the following sequence for the light pulses, which we denote protocol 1. Atoms in $6sn_gs$ states are first created by applying the $6p$ and Rydberg beams at the two-photon resonance during $\sim 1$ $\mu$s, after which ICE light is shined during typically 4 $\mu$s. Signals are maximized by obtaining the largest Rydberg cloud after setting $w_{Ryd} \simeq 500$ $\mu$m and $w_{6p} \simeq 300$ $\mu$m. In order to strike most of the Rydberg atoms, the ICE beam was collimated with effective waist radius set to $w_{ICE} = 560 \pm 10$ $\mu$m.

To measure the energy shift imposed on the $6sn_gs$ state by ICE light, we performed Rydberg excitation spectroscopy under ICE illumination by varying $f_{Ryd}$ while keeping $f_{ICE}$ constant. The sequence of pulses, denoted protocol 2, now consists of the simultaneous application of the three beams during $3$ to $6$ $\mu$s. To increase the shift, the ICE beam was focused on the Rydberg atom cloud: the measurement campaign was conducted with two different values for $w_{ICE}$, $w_{ICE} =  83$ $\mu$m or $105$ $\mu$m. For a better sensitivity of the signals to the shift, the size of the Rydberg cloud was also reduced by setting $w_{Ryd} \simeq 200$ $\mu$m, and $w_{6p} \simeq 300$ or $80$ $\mu$m. Two different variants of protocol 2 were used depending on whether or not an AI signal was detectable:
\begin{itemize}
	\item Away from the AIz, following a protocol denoted 2.1, ICE light pulses were applied for all pulse sequences. Atoms struck by the stronger ICE intensity, thereby undergoing the larger light shift, will also auto-ionize with higher rates, while unshifted atoms will remain in the Rydberg state. Thus, the shift of the Rydberg excitation resonance frequency is manifested as an asymmetry of the ion signal as compared to the Rydberg signal.
	\item At the AIz, no AI signal is detected and we must compare the Rydberg signal obtained with and without the application of ICE light. Following a protocol denoted 2.2, in order to avoid detrimental effects of the thermal drifts of the wavemeter, we implemented a shot-to-shot alternation of the application of ICE light.
\end{itemize}

\section{\label{sec:MQDT}MQDT treatment and data analysis}

\subsection{QDT framework and energy mapping}

For a clear presentation of the data recorded with different $n_g$ numbers, it is beneficial to map energies onto effective quantum numbers in the framework of Quantum Defect Theory (QDT) \cite{Seaton1983}. Effective quantum numbers are of the form $\nu = (n \in  \mathbb{N}^\star) - \delta$, with $n$ the principal quantum number and $\delta$ the quantum defect of the particular Rydberg series considered. We set the origin of energy at the first ionization limit. The energy of a $6sn_gs$ state is given by the Rydberg formula:
\begin{equation} \label{eqn:NRJRydGroundCore}
	E_g=- \frac{hc\mathcal{R}}{\nu_g^2}
\end{equation}
with $h$ the Planck's constant, $c$ the speed of light, $\mathcal{R}$ the Rydberg constant adapted to the Yb atomic mass, and $\nu_g$ is the effective quantum number of the $6sn_gs$ states, explicitly $\nu_g=n_g-\delta_g$. Previous measurements of our team \cite{Lehec2018} yielded accurate values for the quantum defect $\delta_g$. Similarly, the energy $E_e$ in the $6pn_es$ channel is mapped onto $\nu_e = n_e -\delta_e$ ($\delta_e$ is \textit{a priori} unknown) from the $6p_{1/2}$ excited core ionization limit following the relashionship $E_e = hf_{ion} - hc\mathcal{R}/\nu_e^2$. The ICE laser frequency is also mapped onto the effective quantum number $\nu_{ICE}$ in the $6pn_es$ series by equating the addressed energy, $E_{ICE} = E_g +hf_{ICE}$, with $hf_{ion} - hc\mathcal{R}/\nu_{ICE}^2$, leading to:
\begin{equation} \label{eqn:NuICE}
	\nu_{ICE} = \sqrt{\frac{c\mathcal{R}}{-\Delta f_{ICE} + \frac{c \mathcal{R}}{\nu_g^2}}}
\end{equation}
with $\Delta f_{ICE} = f_{ICE} - f_{ion}$ the frequency detuning from the ionic resonance. We also define $\Delta \nu_{ICE}=\nu_{ICE}-\nu_g$; $\Delta \nu_{ICE}=0$ thus corresponds to the ionic core resonance $f_{ICE} = f_{ion}$, while $\Delta \nu_{ICE}=-\left( \delta_e - \delta_g \right)$ corresponds to the \textit{AI resonance}, when the ICE laser addresses resonantly the transition $6sn_gs  \leftrightarrow 6pn_es$ with $n_g = n_e$.

\subsection{MQDT treatment \label{subsec:MQDTtreatment}}

We consider an atom in a $6s n_gs$ state exposed to ICE light with intensity $I$ for a duration $T_\text{Pulse}$. We neglect the direct PI of the Rydberg electron which has negligible probability with our experimental parameters. The atom will experience AI at a rate $\Gamma_{AI} \propto I$ which is evaluated with the AI cross-section $\sigma_{AI}$, defined as follows:
\begin{equation}\label{eqn:DefSigmaAI}
\Gamma_{AI} \left(E_{ICE},E_g\right) = \sigma_{AI} \left(\nu_{ICE}, \nu_g \right) \frac{I}{\hbar \omega_{ion}}
\end{equation}
where $\Phi = I / \left(\hbar \omega_{ion}\right)$ is the photon flux, yielding the dimension of a surface for $\sigma_{AI}$. At the end of the pulse, the probability for its ionization is given by the (Rydberg$\rightarrow$Ion) transfer coefficient:
\begin{equation} \label{eqn:DefTransfer}
	\mathcal{T}_{AI} = 1 - e^{-\sigma_{AI}\Phi T_\text{Pulse}}
\end{equation}
Although the doubly excited $6pn_e s$ channel couples to two independant continua (see Fig. \ref{fig:YbExperiment}), it was demonstrated that a single effective continuum can be considered \cite{Cooke1985}. We thus use a two-channel MQDT model which predicts an AI cross-section $\sigma_{AI} = \sigma_{MQDT}$ as follows \cite{Seaton1983,Bhatti1983,Cooke1985}:
\begin{align}
	\sigma_{MQDT}(\nu_e,\nu_g) &= \frac{\pi \omega_{ion} d^2}{\epsilon_0 c E_H} A_{e}^2 \left(\nu_e,R^\prime,\delta_e\right) f\left(\nu_e,\nu_{g}\right) \label{eqn:sigmaAI}\\
	\text{with: } f\left(\nu_e,\nu_g\right) &= \nu_e^3 \frac{4 \nu_e \nu_g}{\left(\nu_e + \nu_g\right)^2} \left[\frac{\sin\left[\pi\left(\nu_e - \nu_g\right)\right]}{\pi\left(\nu_e - \nu_g\right)}\right]^2 \label{eqn:Deff}\\
	A_e \left(\nu_e,R^\prime,\delta_e\right)  &= -R^\prime \sqrt{\frac{1+\tan^2\left[\pi \left(\nu_e + \delta_e \right)\right]}{{R^\prime}^4 + \tan^2\left[\pi \left(\nu_e + \delta_e \right)\right]}} \label{eqn:Ae}
\end{align}
with $\epsilon_0$ the dielectric constant and $E_H$ the Hartree energy. $R^\prime$ and $\delta_e$ are the MQDT parameters depending on the particular doubly excited channel considered. The function $f\left(\nu_e,\nu_{g}\right)$ includes the overlap integral for the Rydberg electrons \cite{Bhatti1981,Tran1982} which captures in $\sigma_{MQDT}$ the AIz located at ICE frequencies $\Delta \nu_{ICE} \in \mathbb{N}^\star$. This expression assumes that $\delta_e$ is constant over the $6pn_e s$ channel, which is valid far from the perturbation due to any other bound channel \cite{Cooke1985}. In Eq. \eqref{eqn:sigmaAI}, within the ICE framework, $d$ is the modulus of the (vectorial) electric dipole coupling term between $6s$ and $6p_{1/2}$ states of the Yb$^+$ ion; explicitely, with linear polarization of the ICE electric field parallel with $\left( Oz \right)$ axis taken as quantization axis, we have $d = \left| \left< 6s_{1/2} ,m_J | -e z |6p_{1/2}, m_J \right> \right|$ with $e$ the electron charge, $z$ the projection of position operator $\mathbf{r}$ along $\left(Oz\right)$, and $m_J = \pm 1/2$ the quantum number for the projection of the total angular momentum along $\left(Oz\right)$. To evaluate its value, we use the measurements of the natural linewidth of the Yb$^+$ $6p_{1/2}$ state, $\Gamma_{sp} = 2\pi \cdot 19.6$ MHz, and the Weisskopf-Wigner formula for the radiative decay of a two-level atom \cite{Cohen1992}:
\begin{equation} \label{eqn:CoreDipole}
	d=\frac{1}{\sqrt{3}}\sqrt{\Gamma_{sp} \cdot \frac{3 \pi \epsilon_0 \hbar c^3}{{\omega_{ion}}^3}} \simeq 1.01 ea_0
\end{equation}
with $a_0 \simeq 52.9$ pm the Bohr radius. The $1/\sqrt{3}$ prefactor in Eq. \eqref{eqn:CoreDipole} comes from geometrical factors linked to the addressed transition of the multilevel ion.

Eq. \eqref{eqn:Deff} provides a practical scaling law with $n_g$ for the AI cross-section, because it shows that $\sigma_{MQDT}\left(\nu_{ICE},\nu_{g} \right)/\nu_{ICE}^3$ as a function of $\Delta \nu_{ICE}$ is independent of $\nu_g \sim n_g >> \Delta \nu_{ICE}$. Therefore, to display AI cross-sections measured for different $n_g$ numbers and compare it with the MQDT predictions, we will rather rely on the rescaled cross-sections $\sigma_{AI}/\nu_{ICE}^3$.

According to the well-known generic treatment for systems exhibiting discrete excited states which relax to a continuum \cite{Fano1961,Cohen1992}, the energy of the $6s n_gs$ Rydberg atom under ICE excitation will also be shifted by an amount $\left(\Delta E\right)_{ICE}$, linked to the AI rate by a relationship of the Kramers-Kronig type:
\begin{equation} \label{eqn:ShiftICEIntNue0}
	\left(\Delta E\right)_{ICE} = \mathcal{P} \left\{ \int_0^{+\infty} d E_e \frac{\hbar}{2 \pi} \frac{\Gamma_{AI}\left(E_e, E_g \right) }{E_{ICE} - E_e}  \right\}
\end{equation}
\noindent where $\mathcal{P}$ indicates the Cauchy principal value of the integral. Insertion of Eq. \eqref{eqn:sigmaAI} into Eq. \eqref{eqn:DefSigmaAI}, and the latter into Eq. \eqref{eqn:ShiftICEIntNue0}, yields the following shift per unit intensity $\left(\Delta E\right)_{ICE} / I$ as a function of $\nu_{ICE}$:
\begin{equation} \label{eqn:ShiftICEIntNue}
	\frac{\left(\Delta E\right)_{ICE}}{I} = \frac{d^2}{2 \epsilon_0 c E_H}\,\,\mathcal{P} \left\{ \int_0^{+\infty} d \nu_e \frac{g\left(\nu_e\right)\,A_e^2}{\nu_{ICE} - \nu_e} \right\}
\end{equation}
\begin{equation} \label{eqn:Defg}
	\begin{split}
		\text{with:}\,\, g\left(\nu_e,\nu_g,\nu_{ICE} \right) = &\frac{8 \nu_g \nu_e^3 \nu_{ICE}^2}{\left(\nu_e + \nu_{ICE}\right) \left(\nu_g + \nu_e \right)^2} \\
		&\times \left[\frac{\sin\left[\pi\left(\nu_e - \nu_g\right)\right]}{\pi\left(\nu_e - \nu_g\right)}\right]^2
	\end{split}
\end{equation}

\subsection{Data analysis} \label{subsec:FitModels}

For the best evaluation of AI rates and energy shifts imparted by ICE, the spectra recorded were fitted using simulations of the signals which account for spatial variations in the Rydberg cloud density $n_R$ and in the ICE light intensity $I$. We assumed that the ICE beam is well-centered on the Rydberg cloud. After integrating $n_R$ along the ICE propagation axis, we consider 2D Gaussian axisymmetric distributions for $I \left(r \right)$ with waist radius $w_{ICE}$, and for surface density $n_R \left(r \right)$, introducing an effective waist radius $w_R$ to describe the Rydberg cloud size; physically, $w_R$ depends on the properties of the $6p$ and Rydberg beams (see Sec. \ref{sec:Experiment}). Note, that generally $w_{R} \ne w_{Ryd}$, and importantly, that $n_R\left(r \right)$ denotes explicitely the Rydberg atom density when the cloud is created at the two-photon resonance, and without the application of ICE light: Detuning $f_{Ryd}$ from excitation resonance, and/or shifting the latter with ICE light, precisely modifies the rate of Rydberg excitation. The application of ICE light with intensity $I\left(r\right)$ will also possibly ionize Rydberg atoms with cross-section $\sigma_{AI}$. Overall ion and Rydberg counts are then computed by surface integration.

\noindent When protocol 1 is used for AI spectra measurements, i.e. when ICE light is shined after the Rydberg cloud is created with surface density $n_R\left(r \right)$, such integration yields analytical expressions for the transfer coefficient in the form of incomplete Gamma functions.

\noindent When ICE light is shined during Rydberg excitation such as for protocol 2.1, a supplementary integration over time is needed. To put figures, we consider simultaneous pulses starting at $t=0$ and ending at $t=T_\text{Pulse}$. We derive the total surface density of Rydberg excitations by time-integrating the constant rate provided by the product of $n_R\left(r\right)/T_\text{Pulse}$ and a Rydberg state excitation profile depending on $f_{Ryd}$ (of maximal value 1 at resonance), with resonance shifted by $h K_S I\left(r\right)$ where $K_S$ is the shift per unit intensity $\left(\Delta E\right)_{ICE} / I$; subsequent surface integration yields $N_\text{Tot}$. The ion count $N_\text{Ion}$ is derived with another integration over time, a Rydberg atom created at time $\tau$ undergoing an auto-ionizing ICE light pulse of duration $T_\text{Pulse}-\tau$. The total number of Rydberg excitations which survived the ICE pulse is finally deduced following $N_R = N_\text{Tot} -  N_\text{Ion}$. The two surface integrations needed to simulate $N_\text{Tot}$ and $N_R$ are computed numerically and implemented into a dedicated fitting procedure. Note, that in the absence of AI, $N_\text{Tot}$ as compared to the surface integration of $n_R$ simply rescaled by the excitation profile, directly provides the comparison between the two Rydberg signals when following protocol 2.2 at the AIz.

\section{\label{sec:AIspectra}AI spectra}

Our AI spectra measurements are very similar to the pioneering investigations of ICE using atomic beams \cite{Cooke1978b,Cooke1978,Safinya1979,Bhatti1981,Tran1982,Bhatti1983,Jopson1983,Xu1986}. Here, effort was directed towards the frequency resolution of the spectra, the quantitative evaluation of the AI cross-section, and its reduction near the AIz. We measured the overall ionization transfer coefficient $\mathcal{T}_\text{Exp}=N_\text{Ion}/\left(N_\text{Ion}+N_R\right)$ as a function of the ICE laser detuning $\Delta \nu_{ICE}$ for principal quantum numbers $n_g=$ 50 and 60. We observed deviations of $\mathcal{T}_\text{Exp}$ as compared to the theoretical transfer rate $\mathcal{T}_{AI}$ due to the following defects of the experiment:
\begin{itemize}
	\item  Photo-ionization by ICE light of atoms excited to the $\left(6s6p\right){^1P_1}$ state introduces a small background ion signal, also detected when the ICE beam alone is used, and is thus treated as an offset. It arises from the small fraction of the MOT atoms excited with our Zeeman slower beam.
	\item  At the highest AI cross-sections, the transfer saturates strongly, but to a value less than 1 by a few percent. This effect is attributed to the Rydberg excitation of a few atoms out of ICE illumination, due, for instance, to speckle in the intensity profiles of the $6p$ and Rydberg beams. This deviation is treated by a saturation parameter, typically fitted to the best value 0.97, applied to the transfer coefficient.
	\item  A more intriguing feature is a narrow peak in the transfer observed close to the ionic core resonant frequency at $\Delta  \nu_{ICE} = 0$. In our fitting procedure, we found that it is best fitted with an extra ionization cross-section of Lorentzian shape centered on $f_{ion}$, with widths 1.1 GHz and 700 MHz respectively for $n_g$=50 and $n_g$=60 (thus much larger than both $\Gamma_{sp}$ and the ICE Rabi frequencies applied). This points towards the dynamical appearance of high-$\ell$ states (with negligible quantum defect) created over time, which we attribute to mixing of their multiplicity with the $6sn_gs$ channel under the influence of the electrostatic fields created by ions.
\end{itemize}
Best fits to the observed transfer coefficient are performed using the simulations presented above and including these experimental defects. Among the fitted parameters, we found Rydberg cloud waist values $w_R \sim 400$ $\mu$m, consistently comparable to the values of $w_{Ryd}$ and $w_{6p}$. The reciprocal function of the incomplete Gamma function being painful to compute without extra insights into the physics involved, we chose to present both the experimental cross-section for overall ionization $\sigma_\text{Exp}$, and its fit $\sigma_\text{Fit}$, using the average photon flux over the Rydberg cloud with the best fit value for $w_R$:
\begin{equation} \label{eqn:PhiR}
	\Phi_R =\frac{2 P_0}{\pi \hbar\,\omega_{ion} \left(w_{ICE}^2 + w_{R}^2 \right)}
\end{equation}
This immediatly yields $\sigma_\text{Exp}$ by assuming $\mathcal{T}_\text{Exp} = 1 - \exp \left(-\sigma_\text{Exp} \Phi_R T_\text{Pulse} \right)$, and $\sigma_\text{Fit}$ following a similar equation which accounts for defects.

\begin{figure}[h]
	\includegraphics[width=0.99\columnwidth]{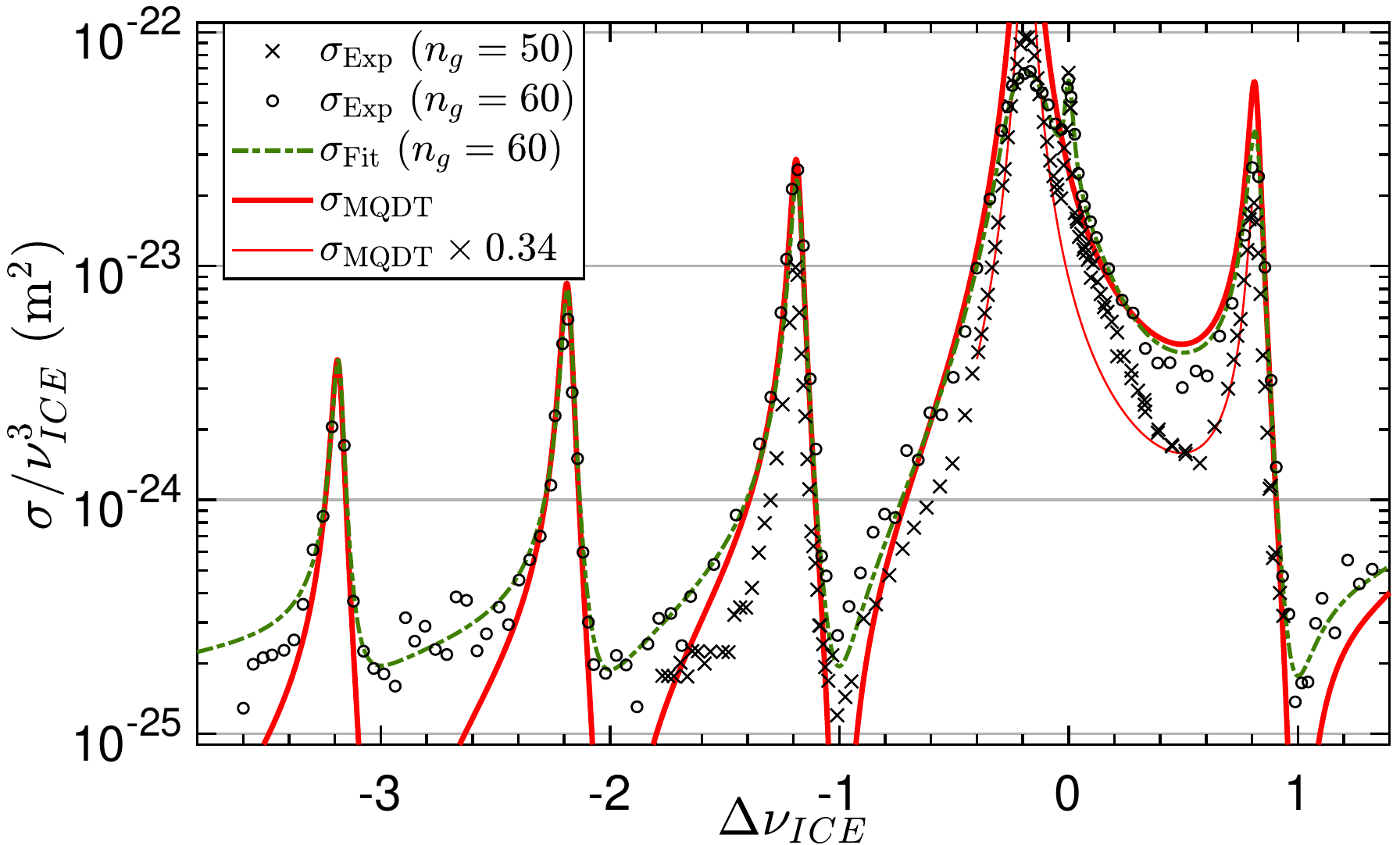}
	\caption{\label{fig:CrossSection}Results of AI spectra measurements, displayed as overall ionization cross-sections $\sigma$ rescaled by $1/\nu_{ICE}^3$ (see Subsec. \ref{subsec:MQDTtreatment}), for $n_g=50$ (circles) and $n_g=60$ (crosses). The results of the fit to the experimental $n_g=60$ transfer coefficient $\mathcal{T}_\text{Exp}$ (dash-dot green line) is plotted together with the theoretical cross section computed from the fitted $R^\prime$ and $\delta_e$ values (thick red plain line) for comparison. The peak value predicted for the AI cross-section at the AI resonance $\Delta \nu_{ICE} \simeq -0.19$ is $\sigma_{MQDT}/\nu_{ICE}^3 = 1.2 \times 10^{-21}$ m$^2$ and corresponds to a saturation parameter $\sigma_\text{MQDT} \Phi_R T_\text{Pulse} \simeq 50$. The $n_g=50$ AI cross-section is found smaller than the theroetical predictions by a factor $\simeq$ 0.34. For comparison, we also plot the rescaled theoretical MQDT cross-section (thin red line). The $n_g=50$ data exhibits an additional pedestal centered at $\Delta \nu_{ICE} = 0$ and interpreted as a Förtser resonance energy transfer.}
\end{figure}

Fig. (\ref{fig:CrossSection}) shows the two measured AI spectra and compares them with the expected theoretical cross section $\sigma_{MQDT}$.
For $n_g=60$ we find excellent agreement between $\sigma_\text{Exp}$ and $\sigma_{MQDT}$ over several orders of magnitude, without any scaling factor for $\sigma_{MQDT}$ (or equivalently the ICE light intensity). For $n_g=50$, we observe a cross-section three times smaller than predicted. In addition, a pedestal in $\sigma_\text{Exp}$, is also visible around the sharp peak at ion resonance. We interpret these specific discrepancies as follows: Inspecting the energies of the $6sn_gs\,^1S_0$ \cite{Lehec2018} and the $6snp\,^{1,3}P_1$ \cite{Aymar1984} Rydberg series, we find that the $6s50s\,^1S_0$ state is in F\"orster resonance \cite{Faoro2015}, within the known uncertainties, with a pair of $6snp\,^{1,3}P_1$ states, leading to high interactions evaluated between 1 to 10 MHz at our density. Moreover, the observed pedestal is compatible with the autoionization of the $6snp\,^{1,3}P_1$ states (see Fig. 4.13 in \cite{Lehec2017}). It thus seems that a significant number of $6s50s\,^1S_0$ atoms turn into $6snp\,^{1,3}P_1$ states which auto-ionize at different ICE laser detunings. In the regions of larger AI rates, this phenomenon is also expected to exalt the mixing with high-$\ell$ states, explaining the reduction of the observed AI cross-section as compared to the predictions. We account for this effect by fitting a rescaling coefficient for the AI cross section, finding a best value $\sigma_{AI} \simeq 0.34 \times \sigma_{MQDT}$, and a Gaussian effective supplementary cross-section for the pedestal, to recover an excellent agreement between the simulated value $\sigma_\text{Fit}$ (not shown in fig. \ref{fig:CrossSection}), and $\sigma_\text{Exp}$ for $n_g=50$. To test the consistency of this explanation, we checked that a relaxation of $6sn_gs$ states to non-autoionizing states with a characteristic time $\sim 3$\,$\mu$s (characteristic frequency of $\sim 50$ kHz) is enough to explain a 0.34 factor in the visible cross section.

The parameters $R^\prime$ and $\delta_e$ (modulo 1) were found to be respectively close to 0.28 and 0.462, with less than $10^{-2}$ relative difference between the cases $n_g=50$ and $n_g=60$. Their stability excludes the presence of an additional strongly perturbing channel and validates the 2-channel approach. One can also observe visible multiple resonances, known as shake-up satellites \cite{Jopson1983}, and corresponding to the resonant excitation of $6pn_es$ states of different principal quantum number $n_e \ne n_g$. They are relatively strong and will hinder the state's coherence over optical manipulation if the ICE laser is mistakingly set to one of them. Most importantly, we confirmed the presence of AIz located at integral values of $\Delta \nu_{ICE}$. At the AIz, the residual ion signal visible in Fig \ref{fig:CrossSection} was found equal to the ion signal measured when $6p$ and Rydberg beams are shut off: whithin our experimental detectivity, this confirms the total cancellation of AI.

\section{ICE light shift}

To test for the existence of a shift imparted by ICE light and compare it to the value predicted by the MQDT model, we performed Rydberg excitation spectroscopy under ICE illumination for principal quantum numbers $n_g$ = 50, 60 and 72. Note, that the ground and intermediate $^1P_1$ states undergo negligible shift under our experimental conditions. Overall, Rydberg excitation spectra were recorded for a range of effective quantum numbers $-1.2 \lesssim \Delta \nu_{ICE} \lesssim 1.0$. Inspection of Eq. \eqref{eqn:ShiftICEIntNue} shows that the detuning from the AI resonance, rather than from ion resonance, is significant in the evaluation of the shift. We thus denote $f_{AI}$ the frequency of the main AI peak (which depends on $n_g$), i.e. the ICE light frequency satisfying $\Delta \nu_{ICE} = - \left( \delta_e - \delta_g \right) \simeq -0.19$ (see Fig \ref{fig:CrossSection}), and $\Delta f_{AI} = f_{ICE} - f_{AI}$ the detuning from AI resonance. To put numbers, $f_{ion} - f_{AI} \simeq 13.1$ GHz for $n_g = 50$, $7.0$ GHz for $n_g = 60$, and $3.8$ GHz for $n_g  = 72$.

\subsection{Rydberg excitation spectra with AI}

\begin{figure}[h]
	\includegraphics[width=0.99\columnwidth]{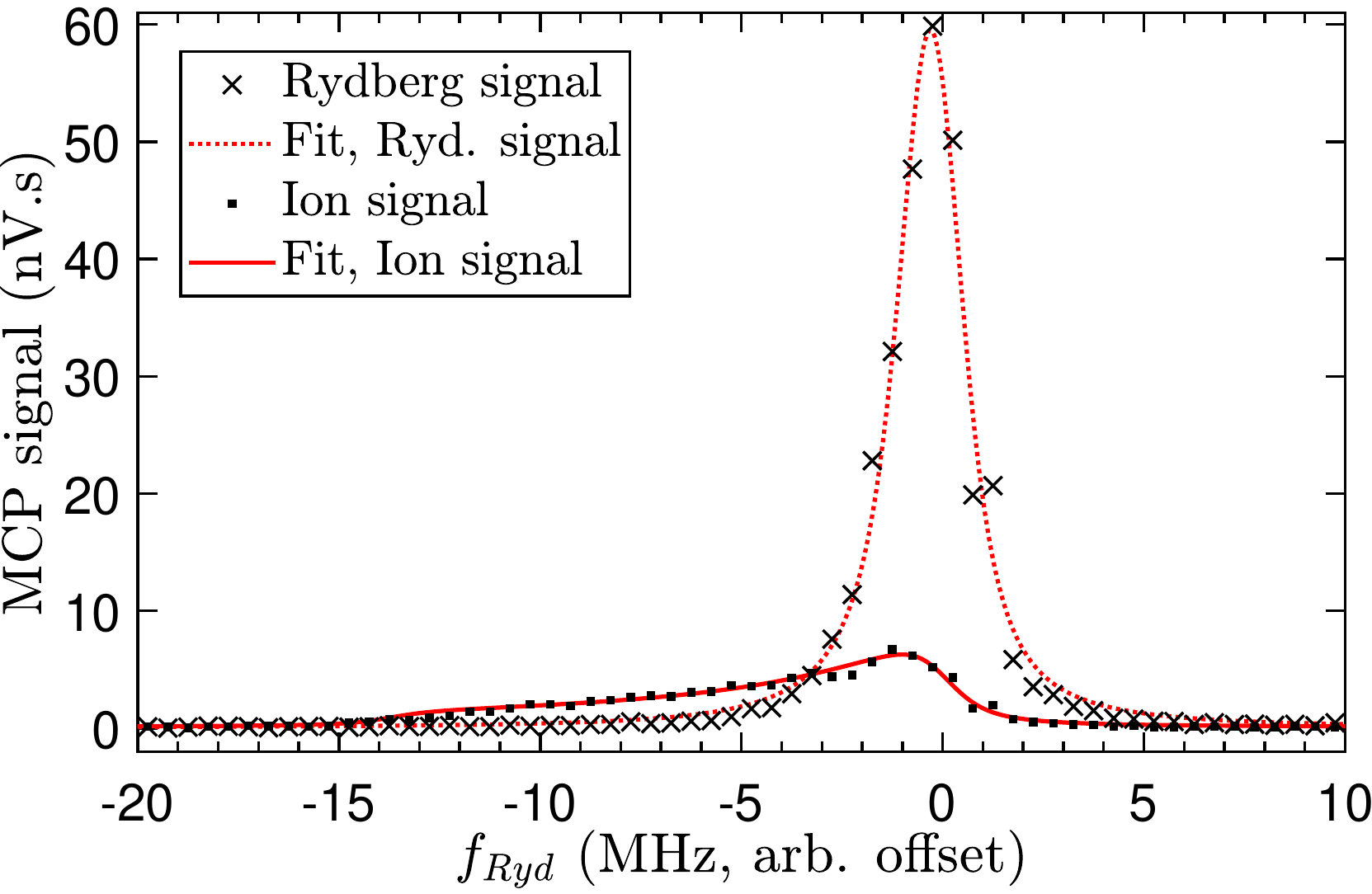}
	\caption{\label{fig:Shift0802Measure4}Rydberg excitation spectroscopy with $n_g = 50$ and ICE laser tuned to $\Delta \nu_{ICE} \simeq -0.365$. Recorded Rydberg (crosses) and ion (squares) signals as a function of the Rydberg excitation laser frequency. The corresponding simulation (dotted and plain lines) use the AI cross section $\sigma_{AI} \simeq 6.14\times 10^{-19}$ m$^2$ and the applied ICE total power $P_{ICE} = 43$ mW.}
\end{figure}

Rydberg excitation spectroscopy out of the AIz was performed with experimental parameters $w_{ICE} = 83$ $\mu$m, $w_{Ryd} \simeq 200$ $\mu$m, $w_{6p} \simeq 300$ $\mu$m, and the pulse sequence of protocol 2.1 with $T_\text{Pulse} = 3$ $\mu$s (see Sec. \ref{sec:Experiment}). For most of these measurements, the best fit value for $w_R$ consistently decreased, as compared to the results of the AI spectra analysis, down to $w_R \simeq 250$ $\mu$m. Simulations of the signals implemented in the data fitting procedure (Subsec. \ref{subsec:FitModels}) used $\sigma_{AI}$ as an input parameter, with value determined by the AI spectra analysis. Thus, for $n_g = 60$ we use $\sigma_{AI}=\sigma_{MQDT}$, while for $n_g=50$ we used $\sigma_{AI} \simeq 0.34 \times \sigma_{MQDT}$ as a best guess for the actual effective AI cross-section. For $n_g= 72$ we did not record any AI spectrum, but we checked the absence of any F\"orster resonance between Rydberg states such as for $n_g  =  50$, and accordingly assumed $\sigma_{AI}=\sigma_{MQDT}$. Fig. \ref{fig:Shift0802Measure4} presents one of the measurements obtained with $n_g = 50$ and $\Delta f_{AI} \simeq -12.3$ GHz ($\Delta \nu_{ICE} \simeq -0.365$). Here and in the following text, numerical values for $\left(\Delta E\right)_{ICE}$ will be given formally as $\left(\Delta E\right)_{ICE} / \left(10^6 h \right)$, i.e. in units of MHz, practical for the observation of spectroscopic measurements.

\noindent The asymmetry of the ion signal is obvious. The agreement of the simulations with the signals is excellent, yielding the measured shift $\left(\Delta E\right)_{ICE} / I \simeq \left(-3.10 \pm 0.25\right) \times 10^{-6}$ MHz/(W.m$^{-2}$) (statistical standard error for $1\sigma$ confidence band). The value predicted by the MQDT model is $\left(\Delta E\right)_{ICE} / I \simeq -2.20 \times 10^{-6}$ MHz/(W.m$^{-2}$). The relative $\sim 40 \%$ (or absolute $\sim 3.5$ standard errors) mismatch is attributed, for one part to experimental fluctuations not accounted for by the simulations such as ICE laser power, MOT position and wavemeter measurement stability, for another part to the limitations of the model used to simulate the signals. Indeed, it uses strong assumptions such as continuity of the Rydberg atom density distribution, and constant Rydberg excitation rate. We expect the model to fail in capturing the physics involved and the experimentally observed signals, for stronger ICE beam focalization, stronger light shifts, and higher AI rates. This phenomenon was actually observed for several measurements performed very close to the AI resonance. With the experimental parameters of the measurement of Fig. \ref{fig:Shift0802Measure4} and according to the results of the fit, a Rydberg atom at the center of the ICE beam undergoes a strong trapping force associated to a negative level shift of 12 MHz: this value can be approximately deduced directly by observing the cutoff in the tail of the ion signal.

\subsection{Measurements at the AI zeros}

In the absence of an ion signal, measurements of the ICE light shift at the AIz follow protocol 2.2 with shot-to-shot alternation of the application of ICE light. They involve relatively large detunings $\left|\Delta \nu_{ICE}\right| \geq 1$, with corresponding shifts approaching the limits of detection with our protocol. Therefore we chose the maximal Rydberg principal number of the measurement campaign $n_g = 72$, because a given AIz at $\Delta \nu_{ICE} \in \mathbb{N}^\star$ corresponds to a smaller detuning $\Delta f_{AI}$, thus larger expected shift, when $n_g$ increases. We reduced further the size of the Rydberg cloud by focusing the $6p$ beam, with an estimated waist radius $w_{6p} \simeq 80$ $\mu$m, yielding best fit values $w_R \sim 150$ $\mu$m.  We also detuned its frequency by 1 GHz from the $^1S_0 \leftrightarrow ^1P_1$ resonance in order to reduce both the frequency width of the Rydberg excitation profile and the background ion signal stemming from the direct PI of atoms in $^1P_1$ state. $T_\text{Pulse}$ was extended to 6 $\mu$s to preserve the overall signal. Finally, we set $w_{ICE} = 105$ $\mu$m as a compromize between strongly shifting a small part of the Rydberg cloud (rather beneficial to exalt the ion signal asymmetry out of the AIz), and illuminating the whole cloud but with weaker shift.
\begin{figure}[h]
	\includegraphics[width=0.99\columnwidth]{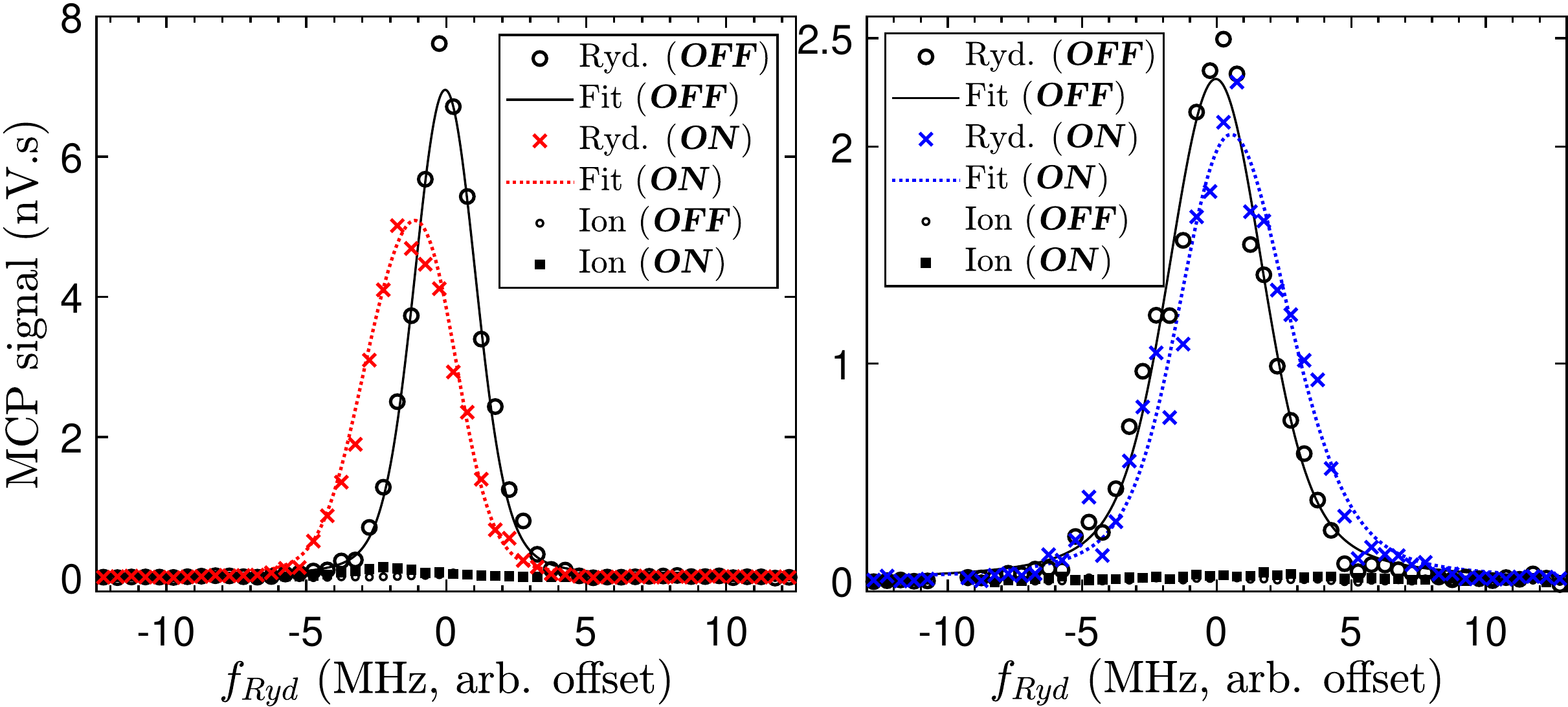}
	\caption{\label{fig:ShiftAIZeros}Evidence for an ICE shift at the AI zeros obtained for $n_g = 72$. \textsl{\textbf{ON}} and \textsl{\textbf{OFF}} denote the application of the $ICE$ beam. The fits (lines) use the signal simulations described in Subsec. \ref{subsec:FitModels} to model both signals simultaneously. Left panel: first red-detuned AI zero ($\Delta \nu_{ICE} = -1$) recorded for $\Delta f_{AI} \simeq -18.1$ GHz with $P_{ICE} = 34$ mW. The relevant results of the best fit to the data are $w_R = \left(128 \pm 2\right)$ $\mu$m, $\left( \Delta E \right)_{ICE} / I = \left(-1.62 \pm 0.04\right) \times 10^{-6}$ MHz/(W.m$^{-2}$), while the predicted shift is $-1.46 \times 10^{-6}$ MHz/(W.m$^{-2}$). Right panel: first blue-detuned AI zero ($\Delta \nu_{ICE} = +1$) recorded for $\Delta f_{AI} \simeq 24.5$ GHz, $P_{ICE} = 29$ mW, resulting in $w_R = \left(200 \pm 15 \right)$ $\mu$m, $\left( \Delta E \right)_{ICE} / I = \left( + 2.1 \pm 0.25 \right) \times 10^{-6}$ MHz/(W.m$^{-2}$) (predicted $+1.5 \times 10^{-6}$ MHz/(W.m$^{-2}$)).}
\end{figure}
Fig. \ref{fig:ShiftAIZeros} shows the signals recorded at the first red and blue-detuned AIz's neighboring the AI resonance ($\Delta \nu_{ICE} = \pm 1$). They show an undeniable shift of the resonance frequency. To our knowledge, no demonstration of an ICE light shift was previously reported, and in particular at an AIz. However, in the spirit of the manipulation of the core electron without disturbing the Rydberg electron, an AIz of a doubly-excited Rydberg intermediate state was utilized to reach higher excited levels of the core with a multiphoton transition \cite{Gallagher1983}.

On the red-detuned AIz (left panel), a small excess ion signal, about 2\% of the Rydberg signal, is visible. For a rough estimate of the finite AI cross-section involved, we use the same procedure as for the extraction of $\sigma_\text{Exp}$ in the AI spectra measurements, rescaling an average photon flux following Eq. \eqref{eqn:PhiR}. With the high intensity imposed and the average ICE pulse duration $T_\text{Pulse}/2 = 3$ $\mu$s, a 2\% transfer coefficient yields $\sigma_\text{Exp} \simeq 4\times 10^{-21}$ m$^2$ or $\sigma_\text{Exp} / \nu_{ICE}^3 \simeq 1.4 \times 10^{-26}$ m$^2$ about 5 orders of magnitude below the resonant AI cross-section (compare Fig. \ref{fig:CrossSection} with $\sigma_\text{MQDT} / \nu_{ICE}^3$ peak at $1.2 \times 10^{-21}$ m$^2$). It is compatible with an inaccurate setting of the laser frequency $\Delta \nu_{ICE} \simeq 0.03$ away from the first, sharp red-detuned AI zero, due to the free running operation of the ICE Ti:sapphire source (with commonly observed drifts up to several 100 MHz). Locking the ICE laser frequency should readily correct this defect and ensure the absence of auto-ionization allowing, for instance, a strong conservative trapping potential for Rydberg states. On the blue-detuned AIz (right pannel), no excess ion signal is clearly visible. To evaluate a higher bound on the AI cross-section obtained, we consider an excess ion signal of 0.01 nV.s, a typical standard deviation for $N_\text{Ion}$ encountered during these measurements, yielding a maximum transfer coefficient of 0.4\%. The same procedure now yields a reduction of almost 6 orders of magnitude of the AI rate as compared to the AI resonance. We are thus convinced that Fig. \ref{fig:ShiftAIZeros} confirms the possibility to apply a strong ICE light shift to an atom in a Rydberg state without AI. By focusing only 10 mW of ICE light on the Rydberg atom with a waist radius of $10$ $\mu$m, we expect a strong shift of amplitude close to 100 MHz which can be used either to apply a force (e.g. trap in the beam at the red-detuned AIz), or engineer its phase without loss or decoherence.

\subsection{Light shift as a function of the detuning}

\begin{figure}[h]
	\includegraphics[width=0.99\columnwidth]{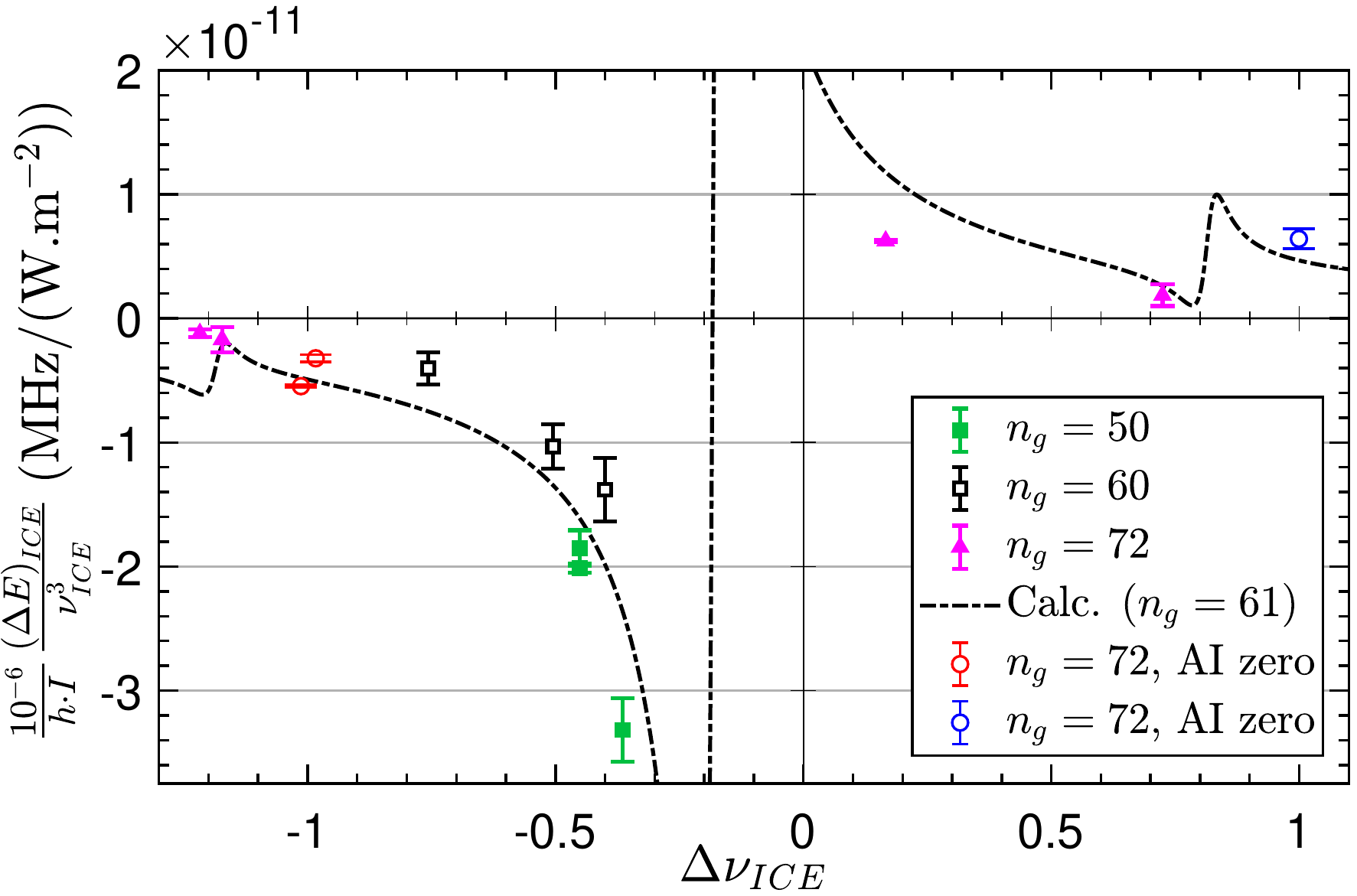}
	\caption{\label{fig:ShiftVsDeltaNu}ICE light shift per unit intensity, rescaled by $1/\nu_{ICE}^3$, extracted from Rydberg excitation spectroscopy at $n_g$ = 50, 60 and 72. The dash-dot line presents the theoretical shift for $n_g = 61$ using Eq. \eqref{eqn:ShiftICEIntNue}.}
\end{figure}

Following Eq. \eqref{eqn:ShiftICEIntNue0}, and with $\sigma_{AI}$ scaling as $\nu_{ICE}^3$ (see Subsec. \ref{subsec:MQDTtreatment}), one can infer the same approximate scaling for $\left(\Delta E\right)_{ICE}$. Therefore, in Fig. \ref{fig:ShiftVsDeltaNu}, we present our results scaled as $\left(\Delta E \right)_{ICE} / \left( \nu_{ICE}^3 10^6 h I \right)$, almost independent of $n_g$, in the same units of MHz/(W.m$^{-2}$). A small theoretical dependence on $n_g$ persists (well within our experimental precision), so we choose to display the theoretical shift for $n_g$=61. We note that the strong shake-up satellite resonances induce local fluctuations in the shift, which are only partially confirmed by our measurements. However, such fluctuations never totally overcome the influence of the main AI feature. Displayed error bars are statistical uncertainties ($1\sigma$ confidence bands) yielded by the data fitting procedure.

We find satisfactory agreement between measured shifts and calculations, and the predicted evolution with $\Delta \nu_{ICE}$ is also well verified. Similarly as for the previous comments on individual shift measurements, for all the data points displayed, the residual disagreement which clearly exceeds statistical uncertainty is well explained by the limitations of the simulations and the level of stability of laser powers, wavemeter measurements and MOT position. Fig. \ref{fig:ShiftVsDeltaNu} thus confirms that the ICE light shift imposed on Rydberg atoms follows the predictions of the MQDT.

\section{Discussion and conclusion}

The AI spectra together with the spectroscopy at the AIz exhibit a dramatic reduction of several orders of magnitude of the AI and the possibility of using close-to-resonance ICE for coherent operations on low-$\ell$ Rydberg states of alkaline-earth species. Experimentally, AIz's are easily accessed, with the requisite frequency accuracy well within the reach of common experimental apparatus. The quasi-resonant character enables the creation of strong shifts which are selectively applied to Rydberg states and drastically reduces light power needed. The short wavelength also increases the spatial resolution when tailoring the ICE intensity distribution.

The minimal AI rate attainable at the AIz's deserves inspection. The two-channel MQDT model predicts no excitation of autoionizing states at the AIz. Accounting for more channels, for example the $6p_{3/2}n_e s$ channel, may lead to non-zero AI rates \cite{Cooke1985}. In close-to-resonance operation, this channel's influence is minimized except for the coincidental existence of a corresponding strong perturbing level in the Rydberg series energy range. The systematic recording of AI spectra for an extended range of $n_g$ numbers should yield important information about possible perturbing channels, which in turn will improve the accuracy of MQDT predictions. Along the same lines, in the prospect of applying ICE to Rydberg states with higher total angular momentum ($J$ quantum number), generally more channels must inevitably be accounted for. Systematic studies with high-$\ell$ states \cite{Jones1988,Pruvost1991,Lehec2021}, but also with excitations of the core above the first optical line \cite{Gallagher1983}, possibly up to the most interesting double Rydberg states \cite{Camus1985,Camus1989,Eichmann1990,Jones1990,Camus1992} should provide valuable insights. These may prove crucial in the engineering of advanced quantum operations with these states, or the investigation of AI with far-off-resonance ICE \cite{Mukherjee2011,Wilson2019} where the consistency of the MQDT model is still to be verified. In the most optimistic case, the decoherence is presumably reduced down to the radiative decay rate of the ionic core, similarly as for circular states.

Close-to-resonance ICE is a new tool offered for experiments aimed at quantum information processing. It will enable the trapping of Rydberg states, which is expected to improve the fidelity of quantum operations \cite{Saffman2016,Madjarov2020}. It will also bring a new degree of freedom for the local addressing \cite{Yavuz2006,Labuhn2014,Xia2015,Wang2015,Wang2016,Graham2019,Levine2019} which leaves the ground-state trapping potential untouched. For instance, one could entangle distant pairs with strong ICE shifts to tune the Rydberg resonances while leaving in-between atoms unperturbed. In this manner, one can expect multiplication of the connectivity of quantum operation \cite{Levine2019,Henriet2020}, or the realization of elaborate quantum gates with more than two qubits.

Choosing the ICE light detuning, one can switch between complementary actions, namely the coherent application of a shift, or the fast AI with a very high certainty. The latter can also be used for the development of technologies for the manipulation of quantum information. At the present time, ICE has only been used to ensure abscence of false detection with optical imaging of ground-state atoms \cite{Madjarov2020}. However, the implementation of a joint detection of atoms by optical means, and single ions with a high detectivity using common electron-multiplying devices, seems within the reach of current apparatus in the near future. Within this framework, implementation of ion optics should strengthen the detection of correlations. Finally, use of AI can constitute a new pathway to engineer samples of ions trapped in optical tweezers with ultimate control \cite{Madjarov2020}.

\section{Acknowledgements}
The authors acknowledge usefull discussions with Eliane Luc-Koenig on the topic of MQDT. This work benefited from financial support by the LabEx PALM (Grant No. ANR-10-LABX-0039). S. L. acknowledges the additional support of CNRS through the program "Tremplin@INP2021".

\end{document}